\documentclass[twocolumn]{elsart3p}
 
\usepackage{amssymb}

\usepackage{graphicx}
\usepackage{amsmath}

\begin{document}

\graphicspath{{./figures/}}

\def\beq{\begin{equation}}
\def\eeq{\end{equation}}
\def\bleq{\begin{eqnarray}}
\def\eleq{\end{eqnarray}} 
\def\bfig{\begin{figure}}
\def\efig{\end{figure}}
\def\bline{\begin{multline}}
\def\eline{\end{multline}}
\def\bremark{\begin{quotation} \noindent \small }
\def\eremark{\end{quotation}}

\newcommand{\Tr}{{\rm Tr}} 
\newcommand{\tr}{{\rm tr}} 
\newcommand{\sgn}{{\rm sgn}} 
\newcommand{\mean}[1]{\langle #1 \rangle}
\newcommand{\commu}[2]{[#1,#2]} 
\newcommand{\ket}[1]{|#1\rangle}
\newcommand{\bra}[1]{\langle#1|}
\newcommand{\braket}[2]{\langle #1|#2\rangle}
\newcommand{\dbraket}[3]{\langle #1|#2|#3\rangle}
\newcommand{\tens}[1]{\overleftrightarrow{#1}}  
\newcommand{\vac}{|{\rm vac}\rangle} 
\def\bravac{\langle{\rm vac}|}
\newcommand{\const}{{\rm const}} 

\newcommand{\ie}{i.e. }
\newcommand{\eg}{e.g. }
\newcommand{\cc}{{\rm c.c.}} 
\newcommand{\hc}{{\rm h.c.}} 

\newcommand{\jhatbf}{\hat {\textbf \j}} 
\newcommand{\Jhatbf}{\hat {\textbf \J}} 
\newcommand{\jhat}{\hat {\jmath}} 
\newcommand{\Jhat}{\hat {J}} 
\newcommand{\jbf}{\textbf j}
\newcommand{\Jbf}{\textbf J}

\def\down{\downarrow}
\def\eps{\epsilon}
\def\gam{\gamma} 
\def\Ome{\Omega}
\def\omeD{{\omega_D}} 
\def\bfOme{\boldsymbol{\Omega}} 
\def\Omebf{\boldsymbol{\Omega}} 
\def\sig{\sigma}
\def\sigp{{\sigma'}} 
\def\bfsig{\boldsymbol{\sigma}} 
\def\sigbf{\boldsymbol{\sigma}} 
\def\The{\Theta} 
\def\up{\uparrow}

\def\epsk{\epsilon_{\bf k}} 
\def\xik{\xi_{\bf k}} 
\def\xikq{\xi_{{\bf k}+{\bf q}}} 
\def\Ek{E_{\bf k}}
\def\Hem{\hat H_{\rm em}}
\def\Hint{\hat H_{\rm int}}
\def\Sem{S_{\rm em}}
\def\SMF{S_{\rm MF}} 
\def\Sint{S_{\rm int}} 
\def\ZMF{Z_{\rm MF}} 
\def\RPA{{\rm RPA}}
\def\loc{{\rm loc}} 
\def\pp{{\rm pp}}
\def\ph{{\rm ph}} 
\def\ch{{\rm ch}}
\def\sp{{\rm sp}} 
\def\qtf{q_{\rm TF}}
\def\epstf{\eps^{}_{\rm TF}} 
\def\epsrpa{\eps^{}_{\rm RPA}} 
\def\chinnzpp{\chi_{nn}^{0}{}\!\!\!''}

\def\half{\frac{1}{2}}
\def\third{\frac{1}{3}} 
\def\quarter{\frac{1}{4}}

\def\qr{{\bf q}\cdot{\bf r}}
\def\wt{\omega t} 

\def\a{{\bf a}}
\def\b{{\bf b}}
\def\f{{\bf f}}
\def\g{{\bf g}}
\def\k{{\bf k}}
\def\l{{\bf l}}
\def\n{{\bf n}} 
\def\p{{\bf p}} 
\def\q{{\bf q}}
\def\r{{\bf r}}
\def\t{{\bf t}}
\def\u{{\bf u}}
\def\v{{\bf v}}
\def\x{{\bf x}}
\def\y{{\bf y}} 
\def\z{{\bf z}} 
\def\A{{\bf A}}
\def\B{{\bf B}}
\def\D{{\bf D}} 
\def\E{{\bf E}} 
\def\F{{\bf F}} 
\def\H{{\bf H}}  
\def\J{{\bf J}}
\def\K{{\bf K}} 

\def\L{{\bf L}}
\def\M{{\bf M}}  
\def\O{{\bf O}} 
\def\P{{\bf P}} 
\def\Q{{\bf Q}} 
\def\R{{\bf R}}
\def\S{{\bf S}}
\def\epsbf{\boldsymbol{\epsilon}}
\def\nablabf{\boldsymbol{\nabla}}
\def\rhobf{\boldsymbol{\rho}}
\def\sigmabf{\boldsymbol{\sigma}} 
\def\Pibf{\boldsymbol{\Pi}}

\def\para{\parallel}
\def\kpara{{k_\parallel}}
\def\kperp{{k_\perp}} 
\def\kperpp{{k_\perp'}} 
\def\qperp{{q_\perp}} 
\def\tperp{{t_\perp}} 

\def\w{\omega}
\def\wn{\omega_n}
\def\wnu{\omega_\nu}
\def\wp{\omega_p} 
\def\dmu{{\partial_\mu}}
\def\dl{{\partial_l}}  
\def\dt{\partial_t}
\def\dx{\partial_x}
\def\dy{\partial_y} 
\def\dtau{{\partial_\tau}}  
\def\det{{\rm det}} 
\def\Pf{{\rm Pf}}

\def\dsum{\displaystyle \sum}
\def\dint{\displaystyle \int} 
\def\intt{\int_{-\infty}^\infty dt} 
\def\inttp{\int_{-\infty}^\infty dt'} 
\def\intk{\int_{\bf k}} 
\def\intkd{\int \frac{d^dk}{(2\pi)^d}}
\def\intq{\int_{\bf q}} 
\def\intr{\int d^dr}  
\def\dintr{\displaystyle \int d^dr} 
\def\intrp{\int d^dr'}
\def\dinttau{\displaystyle \int_0^\beta d\tau}
\def\dinttaup{\displaystyle \int_0^\beta d\tau'}
\def\inttau{\int_0^\beta d\tau}
\def\inttaup{\int_0^\beta d\tau'}
\def\intx{\int d^{d+1}x} 
\def\inttaur{\int_0^\beta d\tau \int d^dr}
\def\intinf{\int_{-\infty}^\infty}
\def\dinttaur{\displaystyle \int_0^\beta d\tau \int d^dr}
\def\dintinf{\displaystyle \int_{-\infty}^\infty}
\def\intw{\int_{-\infty}^\infty \frac{d\w}{2\pi}}
\def\sumr{\sum_{\bf r}} 

\def\calA{{\cal A}} 
\def\calC{{\cal C}}
\def\calD{{\cal D}}
\def\calF{{\cal F}} 
\def\calG{{\cal G}}
\def\calH{{\cal H}}
\def\calJ{{\cal J}}
\def\calL{{\cal L}} 
\def\calN{{\cal N}}
\def\calO{{\cal O}}
\def\calP{{\cal P}}  
\def\calR{{\cal R}} 
\def\calS{{\cal S}}
\def\calT{{\cal T}}
\def\calU{{\cal U}}
\def\calY{{\cal Y}} 

\begin{frontmatter}


\title{Superfluid to Mott-insulator transition of cold atoms in optical lattices}
\author[a1]{N. Dupuis}
and 
\author[a2]{K. Sengupta}
\address[a1]{Laboratoire de Physique Th\'eorique de la Mati\`ere Condens\'ee, CNRS - UMR 7600, \\ Universit\'e Pierre et Marie Curie, 4 Place Jussieu, 75252 Paris Cedex 05, France \\ and \\  Laboratoire de Physique des Solides, CNRS - UMR 8502,\\
  Universit\'e Paris-Sud, 91405 Orsay, France }
\address[a2]{Theoretical Physics Department, Indian Association for the Cultivation of Sciences, Kolkata-700032, India}

\title{}


\author{}

\address{}

\begin{abstract}
We review the superfluid to Mott-insulator transition of cold atoms in optical lattices. The experimental signatures of the transition are discussed and the RPA theory of the Bose-Hubbard model briefly described. We point out that the critical behavior at the transition, as well as the prediction by the RPA theory of a gapped mode (besides the Bogoliubov sound mode) in the superfluid phase, are difficult to understand from the Bogoliubov theory. On the other hand, these findings appear to be intimately connected to the non-trivial infrared behavior of the superfluid phase as recently studied within the non-perturbative renormalization group. 
\end{abstract}

\end{frontmatter}

\section{Introduction}
\label{}

The realization of ultracold atomic gases has opened a new area of research in atomic physics. While early experiments have focused on quantum phenomena associated to coherent matter waves and superfluidity~\cite{revue} (Bose-Einstein condensation, interference between condensates, atom lasers, quantized vortices and vortex lattices, etc.), the emphasis has recently shifted towards strongly correlated systems~\cite{Lewenstein07} following three major experimental developments: the realization of quasi-1D/2D atomic gases using strongly anisotropic confinement traps, the possibility to tune the interaction strength by Feshbach resonances in the atom-atom scattering amplitude~\cite{Kohler06}, and the generation of strong periodic potentials (analog to the crystalline lattice in solids) by optical standing waves. Thus quantum phenomena typical of condensed-matter physics have been observed in cold atomic gases~\cite{Bloch08}.

The Mott transition is one of the most remarkable 
quantum phenomena due to strong correlations. Whereas in solids  
it corresponds to a metal-insulator transition 
driven by the Coulomb repulsion, in bosonic cold atoms in optical lattices it is a transition between a superfluid (SF) and a Mott insulator (MI). This quantum phase transition has been observed in 3D, 1D and 2D ultracold atomic gases~\cite{Greiner02,Gerbier05,Stoferle04,Spielman07}.

\section{The superfluid to Mott-insulator transition in cold atoms}

Ultracold neutral bosonic atoms can be stored in magnetic traps and cooled down to very low temperatures ($\sim 10$ nK) where they condense into a superfluid below a critical temperature $T_c \sim 100 \; \mu {\rm K}$. The importance of the atom-atom interactions can be estimated from the ratio between interaction and kinetic energies, $\gamma \sim n^{1/3}a$, which can be expressed in terms of the particle density $n$ and the $s$-wave scattering length $a$. For Rb atoms in a magnetic trap ($n\sim 10^{14}\; {\rm cm}^{-3}$, $a\sim 5$ nm), 
$\gamma$ is typically of order 0.02, which corresponds to the weakly interacting dilute limit. The physical properties of the gas (collective modes, vortices, etc.) are well described by a macroscopic wave function satisfying the well-known Gross-Pitaevskii equation~\cite{Pitaevskii03}. 

The importance of interactions can be drastically enhanced by subjecting the atomic gas to a periodic lattice potential, which can be created by counter-propagating laser beams. The interference between the two laser beams forms an optical standing wave whose electric field induces a dipole moment in the atom and leads to an interaction energy $V_{\rm OL}(\r)=-\half \alpha(\w_L) |\E(\r)|^2$ with $\E(\r)$ the electric field at position $\r$ ($\alpha(\w_L)$ denotes the polarizability of an atom). By using different arrangement of standing waves, one can create various optical lattices. Three orthogonal standing waves correspond to a 3D cubic lattice and an optical potential
\beq
V_{\rm OL}(\r) = V_0 [ \sin^2(kx) + \sin^2(ky) + \sin^2(kz) ] ,
\eeq
where $k=2\pi/\lambda$ is the wavevector of the laser light. $V_0$ represents the lattice potential depth and is directly related to the laser light intensity. Provided the atoms remain in the lowest Bloch band of the lattice (which requires $V_0$ to be large enough wrt the single atom recoil energy $E_r=\hbar^2 k^2/2m$), the interacting boson gas can be described by the Bose-Hubbard model~\cite{Fischer89,Jaksch98} defined by the Hamiltonian
\begin{multline}
H = - t \sum_{\langle i,j \rangle} (\hat\psi^\dagger_i
\hat\psi_j + {\rm h.c.}) \\ - \sum_i (\mu-\eps_i) \hat n_i  + \frac{U}{2} \sum_i \hat n_i (\hat n_i-1) ,
\label{ham}
\end{multline} 
where $\hat\psi^{(\dagger)}_i$ is a creation/annihilation operator defined at the lattice site $i$, $\hat n_i= \hat\psi^\dagger_i \hat\psi_i$, and $\mu$ denotes the chemical potential fixing the average density of bosons. The on-site energy $\eps_i$ originates form the magnetic trap that confines the atomic cloud as well as the Gaussian shape of the laser beams. The hopping amplitude $t$ and the local repulsion $U$ are given by~\cite{Zwerger03} 
\beq
t = \frac{4E_r}{\sqrt{\pi}} \left(\frac{V_0}{E_r}\right)^{3/4} e^{-2 \sqrt{V_0/E_r}}, 
\eeq
\beq 
U = \sqrt{\frac{8}{\pi}} kaE_r \left(\frac{V_0}{E_r}\right)^{3/4}
\eeq
so that the ratio $t/U$ that controls the physics of the Bose-Hubbard model can be tuned by varying $V_0$, \ie the intensity of the laser beams. 

\begin{figure}
\centerline{\includegraphics[width=6.5cm]{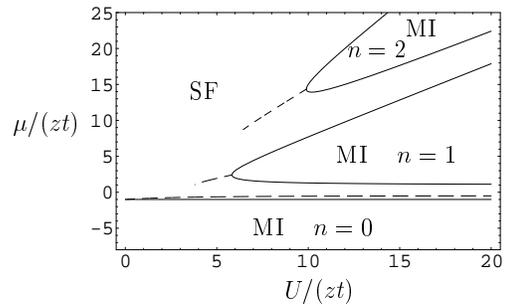}}
\caption{Mean-field phase diagram of the Bose-Hubbard model showing the
  SF phase and the MI phases at
  commensurate filling $n$. The dashed lines correspond to a fixed
  density $n=0.2$, $n=1$ and $n=2$. $z$ denotes the number of nearest neighbors.} 
\label{fig_diagram} 
\end{figure}

In the absence of the confining trap ($\eps_i=\const$), the phase diagram of the Bose-Hubbard model can be obtained from simple arguments. When the kinetic energy dominates ($t/U\gg 1$), the ground state is a superfluid. In the opposite limit of a strong lattice potential ($t/U\ll 1$), the interaction energy dominates and the ground state is a Mott insulator when the average  number of atoms per site is integer. For non-integer fillings, the ground state is always a superfluid irrespective of the strength of the interaction~\cite{Fischer89} (Fig.~\ref{fig_diagram}). The situation is a little more subtle when the confining trap is taken into account~\cite{Jaksch98}. In a local density approximation, the trap potential can be accounted for by a spatially varying chemical potential $\mu_i=\mu-\eps_i$ (taking $\eps_i=0$ at the center of the trap). When moving from the trap center to the edge, the local chemical potential decreases from $\mu$ to zero. We then expect to observe all phases which exist for a chemical potential below $\mu$ (and the same value of $t/U$) in the absence of the trap. If, for instance, the chemical potential $\mu$ falls into the $n=2$ lob in Fig.~\ref{fig_diagram}, the ground state of the gas will correspond to a $n=2$ Mott region at the center of the trap surrounded by a SF region with spatially varying density $1<n<2$, a $n=1$ Mott region and, near the boundary of the trap, a SF region with density $n<1$. The fact that the density remains constant in the MI regions, while the local chemical potential varies, is a consequence of the incompressibility of the Mott phase ($\partial n/\partial\mu=0$).  

The SF-MI transition in cold atoms was first observed by Greiner {\it et al.} in a 3D $^{87}$Rb gas~\cite{Greiner02,Gerbier05}. It has since then also been observed in 1D and 2D gases~\cite{Stoferle04,Spielman07}. The transition is detected by switching off simultaneously the magnetic (confining) and optical potentials and allowing for typically $t=10-20$ ms of free expansion. The density distribution $n(\r,t)$ of the expanding cloud is then obtained by absorption imaging. It can be expressed as~\cite{Kashurnikov02,Gerbier05}
\beq
n(\r,t) = \left(\frac{m}{\hbar t}\right)^3 | \tilde w(\k)|^2 n_\k \Bigl|_{\k=\frac{m\r}{\hbar t}} ,  
\eeq
\beq
n_\k = \sum_{i,j} e^{i\k\cdot (\r_i-\r_j)} \mean{\hat\psi^\dagger_i \hat \psi_j} , 
\eeq
where $\tilde w(\k)$ is the Fourier transform of the Wannier function $w(\r)$ corresponding to the lowest Bloch band of the optical lattice. $n(\r,t)$ therefore provides a direct measure of the momentum distribution $n_\k=\mean{\hat\psi^\dagger(\k) \hat\psi(\k)}$.
In the superfluid state with a finite condensate fraction $f$, the momentum distribution $n_\k$ exhibits sharp peaks -- analogous to the Bragg peaks in the static structure factor of a solid -- at the reciprocal lattice wavevectors $\bf G$. (In a homogeneous superfluid, $f=n_0/n$ with $n_0=\lim_{|\r|\to\infty} \mean{\hat\psi^\dagger_{\r} \hat \psi_0}=|\mean{\hat\psi(\k=0)}|^2/V$ the condensate density.) 
In the MI where $\mean{\hat\psi^\dagger_i \hat \psi_0}$ decays exponentially over a few lattice spacings, the ``Bragg peaks'' are significantly suppressed and broadened~\cite{Zwerger03}. Thus, the superfluid phase is signaled by a high-contrast interference pattern, as expected for a periodic array of coherent matter-wave sources~\cite{Greiner02}. Recent measurements of the momentum distribution function $n_\k$ in the Mott phase of a 2D atomic gas~\cite{Spielman07} (see also~\cite{Gerbier05}) are in excellent agreement (with no adjustable parameter) with the RPA theory~\cite{Sengupta05} discussed in the next section. From the momentum distribution profile, it is also possible to extract a so-called ``coherent fraction'' representing the weight of the sharp peaks and closely related to the condensate fraction $f$, and thus locate the position of the SF-MI transition~\cite{Spielman07}. 

Using an experimental technique based on spatially selective microwave transitions and spin-changing collisions, F\"olling {\it et al.} have directly observed the formation of the spatial shell structure in the SF-MI transition~\cite{Folling06,Campbell06}. This technique enables to record the spatial distribution of lattice sites with different filling factors. As the system evolves from a superfluid to a Mott insulator, it reveals the formation of a distinct shell structure, in agreement with the qualitative discussion given above, and therefore definitively shows the existence of incompressible (Mott) phases. 

Another trademark of the Mott insulator is the existence of a gap in the excitation spectrum whereas the superfluid phase is characterized by a gapless (Bogoliubov) sound mode. Deep in the MI the gap is given by $U$, and should vanish at the MI-SF transition. In the early experiment of Greiner {\it et al.}~\cite{Greiner02}, it was shown that the response of the MI to a potential gradient is compatible with the expected gapped spectrum. In principle, the excitation spectrum can be measured by two-photon Bragg spectroscopy. This technique has allowed to observe the gapless mode of a SF atomic gas~\cite{Steinhauer02}, but for a gas in an optical lattice no convincing result has been obtained so far~\cite{Stoferle04}.

\section{RPA theory... and beyond}

Several theoretical approaches have been used to study the Bose-Hubbard model: mean-field theory~\cite{Fischer89,Sheshadri93,VanOosten01}, random-phase-approximation (RPA)~\cite{Sengupta05,Ohashi06,Konabe06,Menotti08}, strong-coupling expansion \cite{Freericks94}, numerical calculations (Quantum Monte Carlo)~\cite{bosonqmc} 
or variational approach~\cite{Capello07}. In this section, we discuss the RPA theory of the SF-MI transition as well as some open questions related to our understanding of superfluidity. 

\subsection{RPA theory}
\label{subsec_rpa}

The RPA is based on the mean-field decoupling of the hopping term in (\ref{ham}), 
\beq
\hat\psi^\dagger_i \hat\psi_j \to \hat\psi^\dagger_i \phi_j + \phi^*_i \hat\psi_j - \phi^*_i \phi_j 
\eeq
(from now on we ignore the confining trap), where $\phi_i=\mean{\hat\psi_i}$. Taking $\phi_i=\phi_0$ as a uniform order parameter, the Hamiltonian (\ref{ham}) becomes a sum of decoupled one-site Hamiltonians which can be solved exactly~\cite{Sachdev}. A finite value of the order parameter ($\phi_0\neq 0$) signals superfluidity. The resulting phase diagram in the $(U/t,\mu/t)$ plane in shown in Fig.~\ref{fig_diagram}. 

It is possible to go beyond this mean-field approximation in the following way. In the presence of a (fictitious) external source $j_i^*$ that couples to the boson operator $\psi_i$, the (imaginary-time) action reads
\begin{multline}
S_{\RPA} = S_\loc - \inttau \Bigl\lbrace \sum_{i,j} \bigl[ \psi^*_i t_{i,j} \phi_j + \phi^*_i t_{i,j} \psi_j \\ - \phi^*_i t_{i,j} \phi_j \bigr] + \sum_i \bigl[ j^*_i \psi_i + \psi^*_i j_i \bigr] \Bigr\rbrace , 
\end{multline}
with $S_\loc$ the local part ($t=0$) and $\beta=1/T$~\cite{note1}. To eliminate the dependence of the partition function $Z[j^*,j]$ on the external source, it is convenient to perform a Legendre transform to obtain the Gibbs free energy
\begin{equation}
\Gamma[\phi^*,\phi] = \Gamma_\loc[\phi^*,\phi] - \inttau \sum_{i,j} \phi^*_i t_{i,j} \phi_j ,
\label{rpa}
\end{equation}
where $\Gamma_\loc[\phi^*,\phi]$ is the local part ($t=0$). The value of the order parameter $\phi_0=\mean{\psi_i}_{j^*=j=0}$ follows from the equation of state $\delta\Gamma/\delta\phi_i(\tau)|_{\phi_0}=\delta\Gamma/\delta\phi^*_i(\tau)|_{\phi_0}=0$ and determines the condensate density $n_0=|\phi_0|^2$. The Green function is obtained from the second-order functional derivative of (\ref{rpa})~\cite{Negele}. In the RPA, it takes the simple form
\begin{equation}
G^{-1}(\q,i\wn;\phi_0) = G_\loc^{-1}(i\wn;\phi_0) + t(\q) ,
\label{green}
\end{equation}
where $t(\q)$ is the Fourier transform of the hopping amplitude $t_{i,j}$. Equation (\ref{green}) is typical of a strong-coupling expansion in $t/U$ and becomes exact in the limit $t\to 0$~\cite{Freericks94,Sengupta05}. The local Green function $G_\loc(i\wn;\phi_0)$ should be calculated in the presence of an external source $j^*,j$ such that $\mean{\psi}=\phi_0$. Note that in the SF phase $G$ and $G_0$ are $2\times 2$ matrices with both normal (e.g. $\mean{\psi\psi^*}$) and anomalous (e.g. $\mean{\psi\psi}$) components. 

\begin{figure}[h]
\begin{center}
\includegraphics[bb=40 445 280 585,width=7cm]{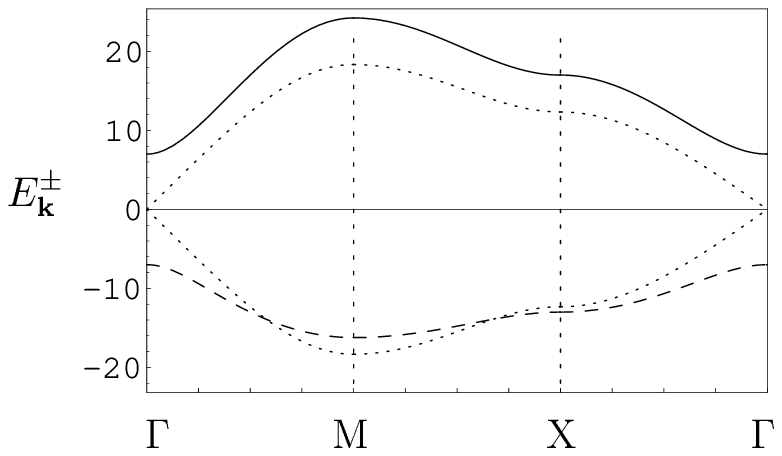}
\includegraphics[bb=40 445 280 585,width=7cm]{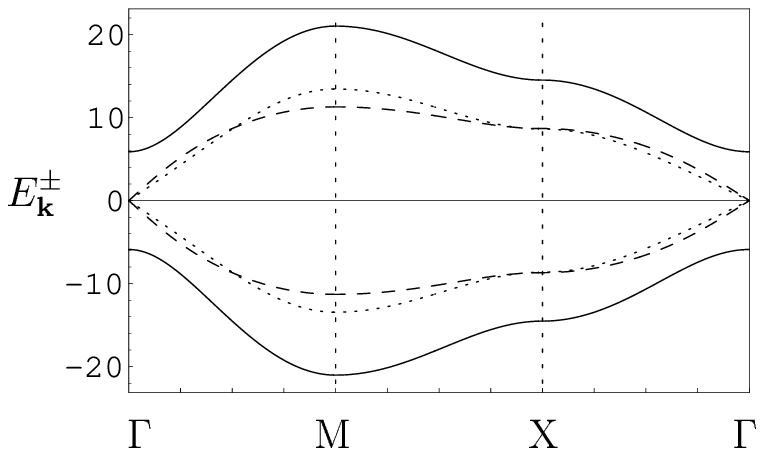}
\end{center}
\caption{Excitation spectrum in 2D. Top: $n=1$ MI 
  ($U/t=30$); bottom:  SF phase ($U/t=20$)~\cite{Sengupta05}. The dotted lines show the result obtained
  from the Bogoliubov theory (which predicts the phase to be
  superfluid).  [$\Gamma=(0,0)$, ${\rm M}=(\pi,\pi)$ 
  and ${\rm X}=(\pi,0)$.] }
\label{fig_modes}
\end{figure}

Equation (\ref{green}) is the central result of the RPA theory of the Bose-Hubbard model. Superfluidity sets in when the boson propagator becomes gapless, \ie $G^{-1}_\loc(i\wn=0)+t(\q=0)=0$. The resulting mean-field phase diagram is qualitatively correct in 3D and 2D (Fig.~\ref{fig_diagram}). Equation (\ref{green}) can also be used to obtain the momentum distribution $n_\k$~\cite{Sengupta05}. In the MI, the agreement with the experimental results is remarkable~\cite{Gerbier05,Spielman07}. The excitation spectrum is obtained from the poles of the Green function $G(\q,\w+i0^+;\phi_0)$. In the MI, one finds two gapped modes as shown in Fig.~\ref{fig_modes} for a 2D system. In the SF phase near the MI-SF transition, there are four excitation modes, two of which being gapless (sound modes) for $\q\to 0$~\cite{Sengupta05,Ohashi06,Konabe06,Menotti08,Huber07}. In the limit $U\to 0$ and for a large number of bosons per site, the two gapped modes disappear and the RPA theory reproduces the result of the Bogoliubov theory~\cite{Menotti08}. These findings suggest that the Bogoliubov theory might not be appropriate, even on a qualitative level, to describe the superfluid phase near the SF-MI transition. 

\subsection{Critical behavior at the SF-MI transition}

The limitations of the Bogoliubov theory are also apparent when one considers the critical behavior at the SF-MI transition. The critical theory can be studied from the action~\cite{Sengupta05,Sachdev}
\begin{multline}
S = \inttau \int d^dr \bigl[Z |\nablabf \psi|^2 + Z_1 \psi^*\dtau \psi \\ + V |\dtau\psi|^2 + r_0 |\psi|^2 + u |\psi|^4 \bigr] .
\label{action}
\end{multline} 
The coefficient $Z_1$ is related to the inverse of the slope of the transition line in Fig.~\ref{fig_diagram} and vanishes at the tip of the Mott lob. The MI-SF transition is then in the universality class of the XY model in $d+1$ dimensions with a dynamic critical exponent $z=1$. It occurs at a fixed density and is accompanied by a vanishing of the excitation gap of the MI. Away from the tip of the Mott lob, $Z_1$ is nonzero and the transition has a dynamic critical exponent $z=2$~\cite{Fischer89}. The density is not conserved and only one of the two modes of the MI becomes gapless (\ie the gap of the MI does not vanish at the transition). Not only does the Bogoliubov theory always predict the system to be superfluid (irrespective of the strength of the interactions), but it also corresponds to $Z_1=1$ and $V=0$ and therefore appears to be a rather poor starting point to understand the SF-MI behavior.  

\subsection{Infrared behavior in the superfluid phase} 

The Bogoliubov theory provides a microscopic explanation of superfluidity and many of its predictions have been confirmed in ultracold atomic gases. Nevertheless a clear understanding of the infrared behavior of interacting bosons at zero temperature has remained a challenging theoretical issue until very recently. Early attempts to go beyond the Bogoliubov theory have revealed a singular perturbation expansion plagued by infrared divergences due to the presence of the Bose-Einstein condensate and the Goldstone mode~\cite{Gavoret64}. These divergences cancel in most physical quantities but lead to a vanishing of the anomalous self-energy $\Sigma_{12}(q)$ in the limit $q=(\q,\w)\to 0$ although the linear spectrum and therefore the superfluidity are preserved~\cite{Nepomnyashchii75}. This observation seriously calls into question the validity of the Bogoliubov theory where the linear spectrum relies on a finite value of $\Sigma_{12}$ ($\Sigma_{12}(q)=4\pi an_0/m$). The physical origin of the vanishing of the anomalous self-energy is the divergence of the longitudinal correlation function which is driven by the gapless (transverse) Goldstone mode -- a general phenomenon in systems with a continuous broken symmetry~\cite{Patasinskij74}. The coupling between longitudinal and transverse fluctuations is not taken into account in Gaussian fluctuation theories such as the Bogoliubov theory. 

The infrared behavior of zero-temperature Bose systems is now well understood in the framework of the non-perturbative renormalization group (NPRG)~\cite{Wetterich08,Dupuis07} (see also Ref.~\cite{Pistolesi04}). The NPRG approach is based on an exact flow equation satisfied by the Gibbs free energy $\Gamma[\phi^*,\phi]$ (see Sec.~\ref{subsec_rpa}) as fluctuations are gradually integrated out beyond the Bogoliubov theory. It reveals that the Bogoliubov form of the action is essentially modified in the RG process when the spatial dimension $d\leq 3$. In the RG language, this means that the Bogoliubov fixed point is unstable when $d\leq 3$. The new fixed point is characterized by a ``relativistic'' action, \ie an action of the type (\ref{action}) with $Z_1=0$ and $V\neq 0$ whereas the Bogoliubov fixed point corresponds to $Z_1>0$ and $V=0$. (In practice the Bogoliubov theory remains valid in 3D systems as the RG flow is only logarithmic so that the new fixed point is not accessible in any finite size system.)

It is quite remarkable that a proper treatment of a Bose gas in a continuum model (\ie with no underlying periodic lattice) yields a low-energy action similar to that obtained from the Bose-Hubbard model near the MI-SF transition. In other words, the critical behavior at the MI-SF transition appears to be intimately connected to the non-trivial infrared behavior of the superfluid phase. A generalization of the NPRG technique to the lattice case should shed light on this issue. 

\section{Conclusion}

Ultracold atomic gases allow us to study strongly-correlated systems in an unprecedentedly controlled manner. Not only do we have a detailed microscopic understanding of the Hamiltonian of the system realized in the laboratory, but the microscopic parameters that control the physics can be tuned by varying external fields. Hamiltonians typical of strongly-correlated systems (\eg (Bose-)Hubbard models) can be simulated in cold atomic gases, and new systems -- with no equivalent in condensed-matter physics -- can also be created~\cite{Lewenstein07}. 

A great success in the study of cold atoms has been the observation of the SF-MI transition. In this paper, we have reviewed the main experimental signatures of the SF and MI phases. The RPA theory of the Bose-Hubbard model provides a qualitative and sometimes quantitative description of the system. What has not been observed so far is the excitation spectrum and the critical behavior at the transition. Whether the experimental difficulties and the finite size (as well as the non-uniform density) of the atomic clouds will allow these observations is an open question. 

From a more theoretical side, we have pointed out the difficulty to understand the superfluid phase near the SF-MI transition, as well as the critical behavior at the transition, from the Bogoliubov theory. On the other hand, the critical behavior appears to be intimately connected to the non-trivial infrared behavior of the superfluid phase as recently studied within the non-perturbative renormalization group~\cite{Wetterich08,Dupuis07}. 

\section*{Acknowledgment}

ND would like to thank F. Gerbier for enlightening discussions and a critical reading of the manuscript.




\begin{thebibliography}{00}

\bibitem{revue} For a review see F. Dalfovo, S. Giorgini, L.~P. Pitaevskii, and
  S. Stringari, Rev. Mod. Phys. {\bf 71}, 463 (1999); A.~J. Leggett,
  Rev. Mod. Phys. {\bf 73}, 307 (2001); W. Ketterle and M.~W. Zwierlein, arXiv:0801.2500.

\bibitem{Lewenstein07} 
M. Lewenstein, A. Sanpera, V. Ahufinger, B. Damski, A. Sen De, and U. Sen,
Adv. in Phys. {\bf 56}, 243 (2007). 

\bibitem{Kohler06}
T. Kohler, K. Goral and P.~S. Julienne, Rev. Mod. Phys. {\bf 78}, 1311 (2006). 

\bibitem{Bloch08}
I. Bloch, J. Dalibard, and W. Zwerger, Rev. Mod. Phys. {\bf 80}, 885 (2008). 

\bibitem{Greiner02} M. Greiner, O. Mandel, 
T. Esslinger, T.~W. H\"ansch, and I. Bloch, Nature {\bf 415}, 39 (2002). 

\bibitem{Gerbier05} F. Gerbier, A. Widera, S. F\"olling, O. Mandel, T. Gericke, 
and I. Bloch, Phys. Rev. Lett. {\bf 95}, 050404 (2005); Phys. Rev. A {\bf 72}, 053606 (2005).  

\bibitem{Stoferle04} T. St\"oferle, H. Moritz, 
C. Schori, M. K\"ohl, and T. Esslinger, Phys. Rev. Lett. {\bf 92}, 130403 (2004). 

\bibitem{Spielman07} I.~B. Spielman, W.~D. Philipps, and J.~V. Porto, Phys. Rev. Lett. {\bf 98}, 080404 (2007); Phys. Rev. Lett. {\bf 100}, 120402 (2008). 

\bibitem{Pitaevskii03} See, for instance, L.~P. Pitaevskii and S. Stringari, {\it Bose Einstein Condensation} (Oxford Science, Oxford, 2003). 

\bibitem{Fischer89} M.~P.~A. Fisher, P.~B. Weichman,
G. Grinstein, and D.~S. Fisher, Phys. Rev. B {\bf 40}, 546 (1989).  

\bibitem{Jaksch98} D. Jaksch, C. Bruder, 
J.~I. Cirac, C.~W. Gardiner, and P. Zoller, Phys. Rev. Lett. {\bf 81}, 3108 (1998). 

\bibitem{Zwerger03} W. Zwerger, J. Opt. B {\bf 5}, S9 (2003). 

\bibitem{Kashurnikov02} V.~A. Kashurnikov, N.~V. Prokof'ev, and B.~V. Svistunov, Phys. Rev. A {\bf 66}, 031601(R) (2002). 

\bibitem{Sengupta05} K. Sengupta and N. Dupuis, Phys. Rev. A {\bf 71}, 033629 (2005).

\bibitem{Folling06} S. F\"olling, A. Widera, T. M\"uller, F. Gerbier, and I. Bloch, Phys Rev. Lett. {\bf 97}, 060403 (2006). 

\bibitem{Campbell06} See also 
G.~K. Campbell, J. Mun, M. Boyd, P. Medley, A.~E. Leanhardt, L.~G. Marcassa, D.~E. Pritchard and W. Ketterle, Science {\bf 313}, 649 (2006). 
 
\bibitem{Steinhauer02} J. Steinhauer, R. Ozeri, N. Katz, and N. Davidson, Phys. Rev. Lett. {\bf 88}, 120407 (2002). 

\bibitem{Sheshadri93} K. Sheshadri, H.~R. Krishnamurthy, R. Pandit, and
  T.~V. Ramakrishnan, Europhys. Lett. {\bf 22}, 257 (1993).

\bibitem{VanOosten01} D. Van Oosten, P. van der Straten, and
  H.~T.~C. Stoof, Phys. Rev. A {\bf 63},053601 (2001). 

\bibitem{Ohashi06} Y. Ohashi, M. Kitauri, and H. Matsumoto, Phys. Rev. A {\bf 73}, 033617 (2006).

\bibitem{Konabe06} S. Konabe, T. Nikuni, and M. Nakamura, Phys. Rev. A {\bf 73}, 033621 (2006). 

\bibitem{Menotti08} C. Menotti and N. Trivedi, arXiv:0801.4672. 

\bibitem{Freericks94} J.~K. Freericks and H. Monien, Europhys. Lett.
{\bf 26}, 545 (1994); {\it ibid.} Phys. Rev. B {\bf 53}, 2691
(1996).

\bibitem{bosonqmc} F. Alet and E. S\o rensen, Phys. Rev. B {\bf 70}, 024513 (2004); 
B. Capogrosso-Sansone, S.~G. S\"oyler, N. Prokof'ev, and B. Svistunov, Phys. Rev. A {\bf 77}, 015602 (2008) and references therein.

\bibitem{Capello07} M. Capello, M. Becca, M. Fabrizio, and S. Sorella, Phys. Rev. Lett. {\bf 99}, 056402 (2007). 

\bibitem{Sachdev} S. Sachdev, {\it Quantum Phase Transitions} (Cambridge University Press, Cambridge, England, 1999). 

\bibitem{note1} \note 
Near the MI transition, it is natural to expand the functional $\Gamma$ in powers of $\phi$,
\begin{multline}
\Gamma[\phi^*,\phi] = - \sum_{a,b} \phi^*_a [ G^{-1}_\loc(a,b) + t_{a,b} ] \phi_b \\  
+ \quarter \sum_{a,b,c,d} \Gamma^{\rm II}_\loc (a,b,c,d) \phi^*_a \phi^*_b \phi_c \phi_d \nonumber 
\label{gl}
\end{multline}
($a\equiv(\r_i,\tau)$ and $\sum_a\equiv\inttau \sum_i$), where $G_\loc$ is the local propagator for $j^*=j=0$ and $\Gamma^{\rm II}_\loc$ the local two-particle vertex. The preceding equation was obtained in Ref.~\cite{Sengupta05} from a strong-coupling expansion. For a static and uniform field $\phi_\r(\tau)=\phi_0$, it gives the expansion of the Gibbs free energy to $\calO(|\phi_0^4|)$.

\bibitem{Negele} See, for instance, J.~W. Negele and H. Orland, {\it Quantum Many-Particle Systems} (HarperCollins, Canada, 1998). 

\bibitem{Huber07}  S.~D. Huber, E. Altman, H.~P. Büchler, and G. Blatter, Phys. Rev. B {\bf 75}, 085106 (2007).

\bibitem{Gavoret64}
J. Gavoret and P. Nozi\`eres, Ann. Phys. (N.Y.) {\bf 28},  349  (1964).

\bibitem{Nepomnyashchii75}
A.~A. Nepomnyashchii and Y.~A. Nepomnyashchii, JETP Lett. {\bf 21},  1  (1975);
Y.~A. Nepomnyashchii and A.~A. Nepomnyashchii, Sov. Phys. JETP {\bf 48},  493
  (1978);
Y.~A. Nepomnyashchii, Sov. Phys. JETP {\bf 58},  722  (1983).

\bibitem{Patasinskij74}
A.~Z. Patasinskij and V.~L. Pokrovskij, Sov. Phys. JETP {\bf 37},  733  (1973).

\bibitem{Wetterich08} C. Wetterich, Phys. Rev. B {\bf 77}, 064504 (2008).

\bibitem{Dupuis07} N. Dupuis and K. Sengupta, Europhys. Lett. {\bf 80}, 50007 (2007).

\bibitem{Pistolesi04}
F. Pistolesi, C. Castellani, C.~D. Castro, and G.~C. Strinati, Phys. Rev. B
  {\bf 69},  024513  (2004).





\end{thebibliography}
\end{document}